\documentclass{aa} % for a referee version
\usepackage{graphicx}
\usepackage{txfonts}
\usepackage{longtable}
\usepackage{amsmath}
\usepackage{graphicx}
\usepackage{float}
\usepackage{multirow}
\usepackage{subfigure}
\usepackage{rotfloat}

\usepackage[]{hyperref}
\usepackage{breakurl}
\newcommand{\urlwofont}[1]{\urlstyle{same}\url{#1}}
\usepackage{array}

\usepackage{booktabs}
\usepackage{lscape}

\begin{document}

\title{Overall Spectral Properties of Prompt Emissions with Diverse Segments in Swift/BAT Short Gamma-ray Bursts}

\author{X.J.~Li\inst{1}
\and Z.B.~Zhang*\inst{1}
\and K.~Zhang\inst{1}
%\and M.~Rapisarda\inst{3}
%%\and E.~Palazzi\inst{1}
}

%\offprints{L. Amati:amati@tesre.bo.cnr.it}

\institute{School of Physics and Physical Engineering, Qufu Normal University, Qufu 273165, China\\
\email{astrophy0817@163.com}}
\date{Received 2021 March 8; Accepted 2021 October}

\markboth{Connections between precursor, main peak and extended emission of Swift Short Gamma-ray bursts}{}

%   \date{Received September 15, 1996; accepted March 16, 1997}

% \abstract{}{}{}{}{}
% 5 {} token are mandatory

\abstract{Owing to lack of multiple components of  prompt $\gamma$-ray emissions in short gamma-ray bursts (sGRBs), how  these distinct components are correlated still keeps unclear. In this paper,
we investigate the spectral and temporal properties of precursors, main peaks and extended emissions in 26 sGRBs including GRB 170817A.
It is found that peak energies ($E_p$) in each pulse are uncorrelated with the pulse duration ($t_{dur}$). Meanwhile, we find that there is no obvious correlation between peak energy and energy fluence. Interestingly, there is no obvious spectral evolution from earlier precursors to later extended emissions in view of the correlations of $t_{dur}$ with either the $E_p$ or the low energy spectrum index, $\alpha$. A power-law correlation between the average flux ($F_{p}$) and the energy fluence ($S_\gamma$), $log F_p=(0.62\pm0.07) log S_\gamma + (0.27\pm0.07)$, is found to exist in the individual segments instead of mean peaks previously. Furthermore, we also find that the main peaks are on average brighter than the precursors or the extend emissions about one order of magnitude. On the basis of all the above analyses, one can conclude that three emissive
components would share the same radiation mechanisms but they might be dominated by
diverse physical processes. }

\authorrunning{Li et al.}
\titlerunning{Connections between precursor, main peak and extended emission of Swift Short GRBs}
\maketitle

\section{Introduction}

The \emph{Swift} satellite was successfully launched in 2004 November \citep{Gehrels2004} and has detected over 1300 gamma-ray bursts (GRBs) till 2019 November. According to the classification criterion of the $T_{90}$ duration distribution \citep{Kouveliotou1993}, approximately 10\% are short GRBs (sGRBs), with a typical duration of $T_{90}$ $<$ 2s \citep{zhangzb2008,zhangzb2018}. Prompt garmma-ray emissions of GRBs may consist of diverse components, namely precursors, main peaks and extended emissions (EEs), or parts of them, within both long GRBs (lGRBs) and sGRBs \citep{Metzger1974,Koshut1995,Norris2006,Troja2010,Bernardini2013,Hu2014,Lan2018,Lan2020,ZhangBB2018b,Zhong2019,Zhangxiaolu2020,Li2021}.
The precursor reported first in GRB 720427 is a dim peak occurring before the brightest prompt emission of main peaks \citep{Metzger1974}, and the EE as the softer $\gamma$-ray emissions usually following the main peaks after a quiescent period is another important component \citep[e.g.,][]{Lazzati2001,Connaughton2002,Burrows2005}.

Some authors argued that there are no obvious correlations between the precursors and the main peaks \citep{Koshut1995,Lazzati2005,Burlon2008,Burlon2009, Charisi2015}. Some others extracted and compared the temporal and spectral characteristics of EEs with main peaks \citep{Norris2010,Norris2011,Bostanci2013,Kaneko2015,Kagawa2015,Lien2016,Anand2018}. For example, \cite{Zhong2019} extracted 18 sGRB candidates with precursor observed by \emph{Fermi/GBM} and \emph{Swift/BAT}. They found that the average flux of precursor components tends to increase as those of the main peaks. They compared the hard ratio and the cutoff energy \emph{E$_{c}$} between these two emission episodes, suggesting that the main peaks are slightly harder than the precursors. Recently, \cite{Lan2020} identified 26 \emph{Fermi/GBM} sGRBs with early EE similar to GRB 060614. Their results suggested that the sGRBs with EE probably have a similar physical origin. Particularly, they compared the properties of GRB 170817A as the first gravitational-wave associated sGRB with EE \citep{Abbott2017,Goldstein2017}, with other sGRBs with EE and found that there are no significant statistical differences between them.

The prompt GRB emissions are often produced by either the quasi-thermal mechanism from photosphere of a fireball or the synchrotron radiation mechanism from electrons of the Poynting-flux-dominated
jet, respectively    \citep{Katz1994,Meszaros1994,Thompson1994,Rees1994,Daigne1998,ZhangandYan2011,DengandZhang2014,Deng2015,BeniaminiandGiannios2017,
Lazarian2019,Meng2018,Meng2019,Liliang2019b,Ryde2019}.
It is generally accepted that the low-energy photon index, is an indicator for the emission mechanism, using to distinguish
the synchrotron and photosphere emissions \citep{Liliang2019a}. Around 50\% of Swift GRBs are better explained by the black body (BB) spectrum for most X-ray flashes (XRFs) plus X-ray-rich GRBs (XRBs) or the synchrotron radiation mechanism for most classical
GRBs (C-GRBs), correspondingly \citep{Oganesyan2019,Zhangzb2020}.
\cite{Lan2018} systematically studied the spectral and temporal properties of two emission episodes separated by quiescent gaps for 101 \emph{Fermi/GBM} lGRBs. They found similar distribution of peak energy between two emission episodes and suggested that these two share the same physical origin. However, it was found that the thermal component appears in the first emission episode and a transition from the thermal to the non-thermal component may exist in multipulse \emph{Fermi} GRBs in the prompt gamma-ray emission phase \citep{ZhangBB2018b,Liliang2019a,Liliang2019b}.

\cite{ZhangBB2018a} studied the time-resolved spectra in each episode of GRB 160625B detected by Fermi, from precursor to main peaks and to extended emission. They announced a transition from thermal to non-thermal mechanism within a GRB. The indisputable fact is that these investigations were mainly given to long bursts, not including short ones due to absence of sGRBs with multiple components.
Recently, we defined two kinds of double-peaked BATSE sGRBs as M-loose and
M-tight types according to their overlapping ratios between two adjacent main peaks \citep [see Fig. 1 in][] {Li2020}. Then, we examined the temporal properties of the main peaks and the other two components of Swift/BAT sGRBs \citep{Li2021}.
%However, a major issue is the very definition of ``precursor and EE''as there is no obvious objective criterion as yet.
We adopted a united criterion
to search for precursors and EEs as their signals prior and posterior to the main peaks at least S/N > 3 above background \citep[see section 2.2 in][]{Li2021}.
Unfortunately, no such triplets have been reported in single sGRB to date.
Considering the above
controversial results, we generalize the spectral analysis and evolution of the three components by using sGRB samples with any two components instead. In this way, one can simultaneously investigate the time-integrated spectral properties of the three components of sGRBs with single or double main peaks. In addition, we will examine how the spectra evolve from precursor, main peak, to extended emission.
Sample selection and spectra analysis are presented in Section 2. Section 3 displays our temporal and spectral results of these sGRBs. We end with the conclusions and discussions in Section 4.

\section{Data analysis method}
From 2004 December to 2019 July, Swift/BAT had detected 124 sGRBs, of which 26 sGRBs have been selected for this study, including 12 single-peaked sGRBs (SPs), 5 double-peaked sGRBs (DPs), 7 sGRBs with precursor (Pre+sGRBs), and 2 sGRBs with EE (sGRBs+EE). For the DPs, we still divide them into the M-tight types (Mt-DPs) and the M-loose ones (Ml-DPs) as done in our recent works \citep{Li2020,Li2021}. The criteria that we identify a significant precursor or EE pulse can be referred to our recent paper \citep{Li2021}. The standard BAT software (HEADAS 6.26.1) and the latest calibration database (CALDB: 2017-10-16) are used. Refer to BAT analysis threads  \footnote{\url{https://swift.gsfc.nasa.gov/analysis/threads/bat_threads.html/} or \url{https://www.swift.ac.uk/analysis/bat/}}
for the handing process.
\begin{figure}[H]%"[]"中为位置参数，四个参数tbph依次是置顶、置底、浮动、当前位置，，选用的参数优先顺序为h-t-b-p
\centering
   \begin{minipage}[c]{1\textwidth}
  \includegraphics[height=4cm, width=8cm, angle=0]{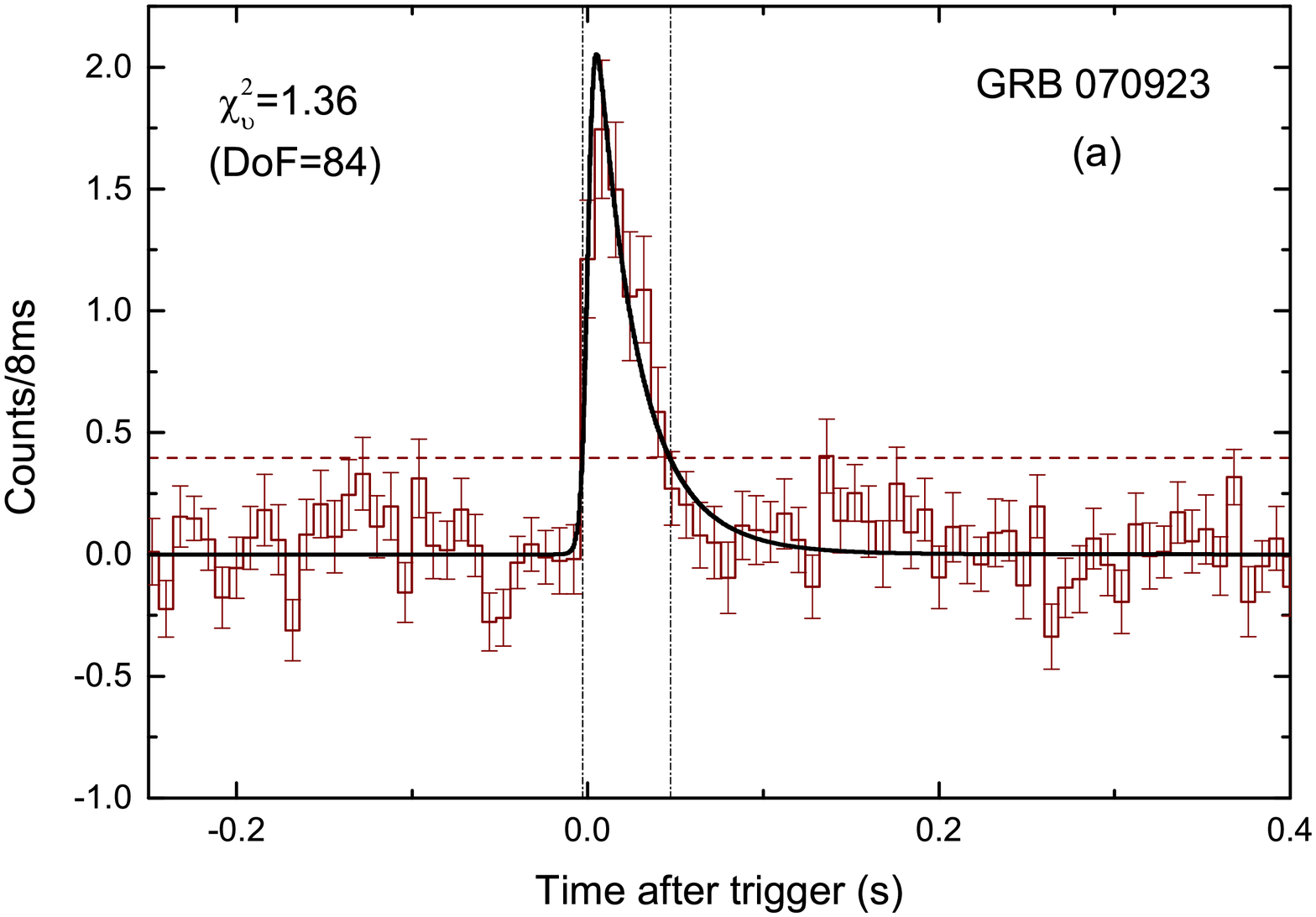}
  \end{minipage}
  \begin{minipage}[c]{1\textwidth}
\includegraphics[height=4cm, width=7.8cm, angle=0]{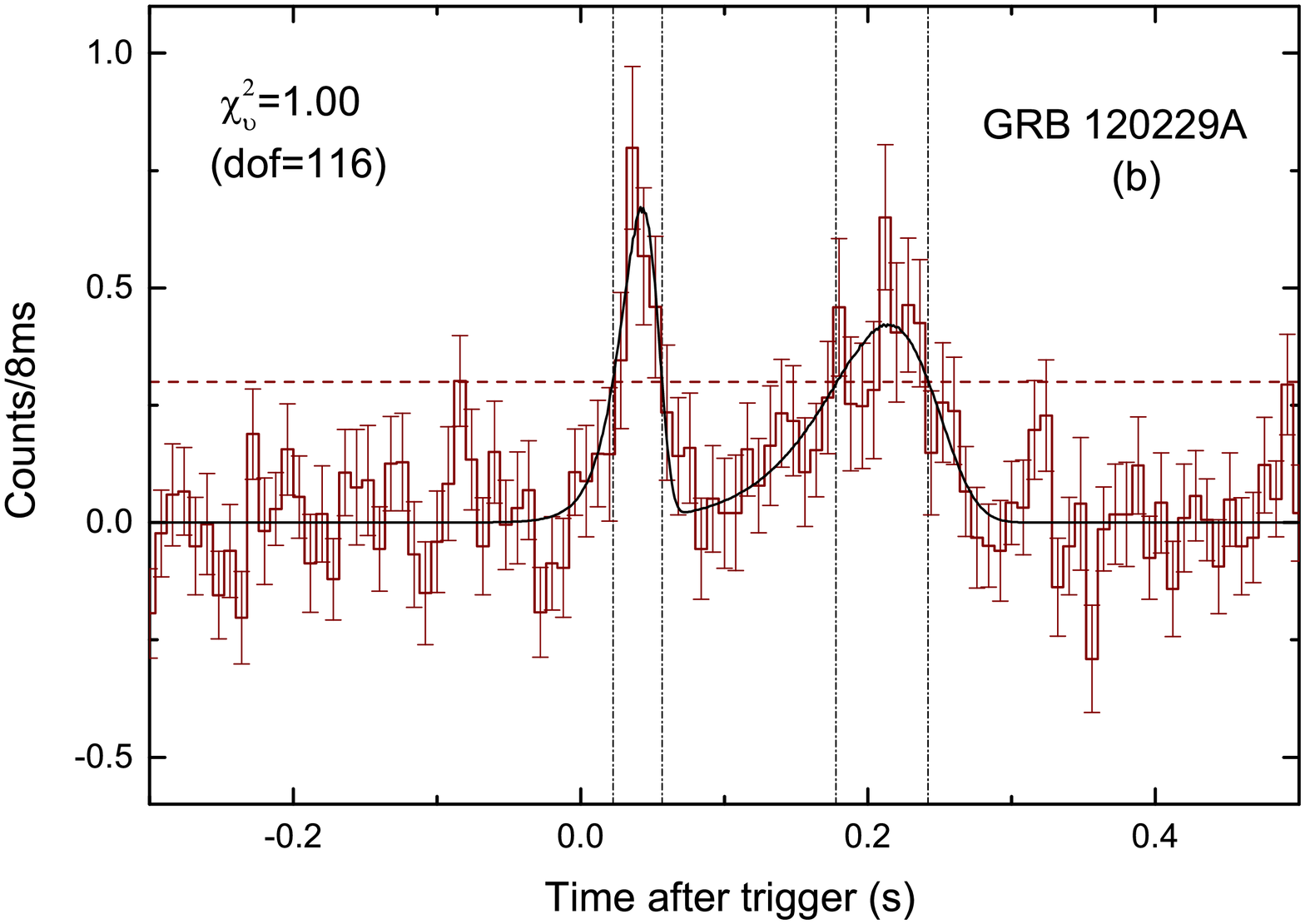}%"scale" 后的数字为图形的宽度，也可用"width=1.0\columnwidth"定义
  \end{minipage}
    \begin{minipage}[c]{1\textwidth}
\includegraphics[height=4cm, width=8cm, angle=0]{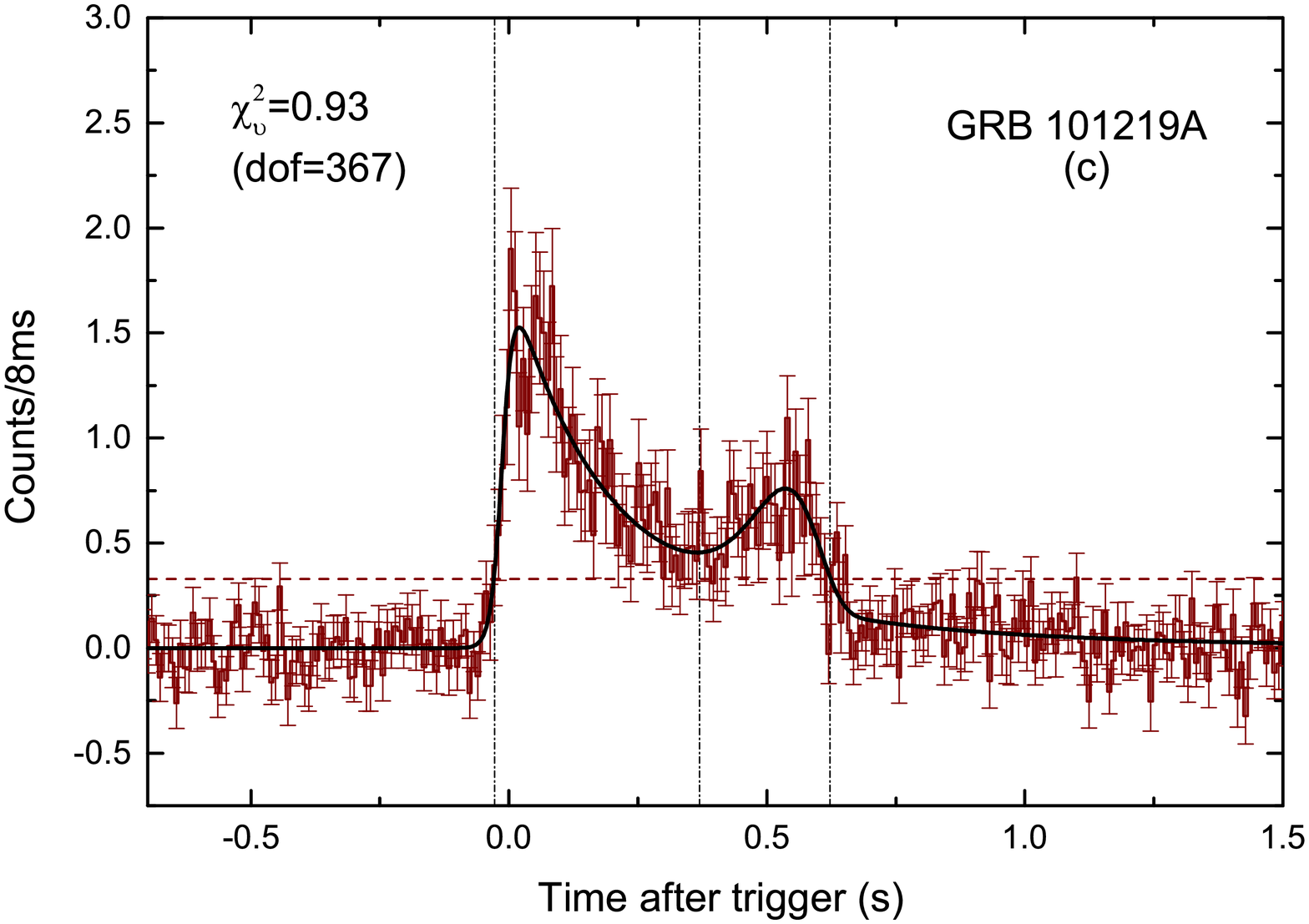}%"scale" 后的数字为图形的宽度，也可用"width=1.0\columnwidth" 定义
  \end{minipage}
    \begin{minipage}[c]{1\textwidth}
\includegraphics[height=4cm, width=8cm, angle=0]{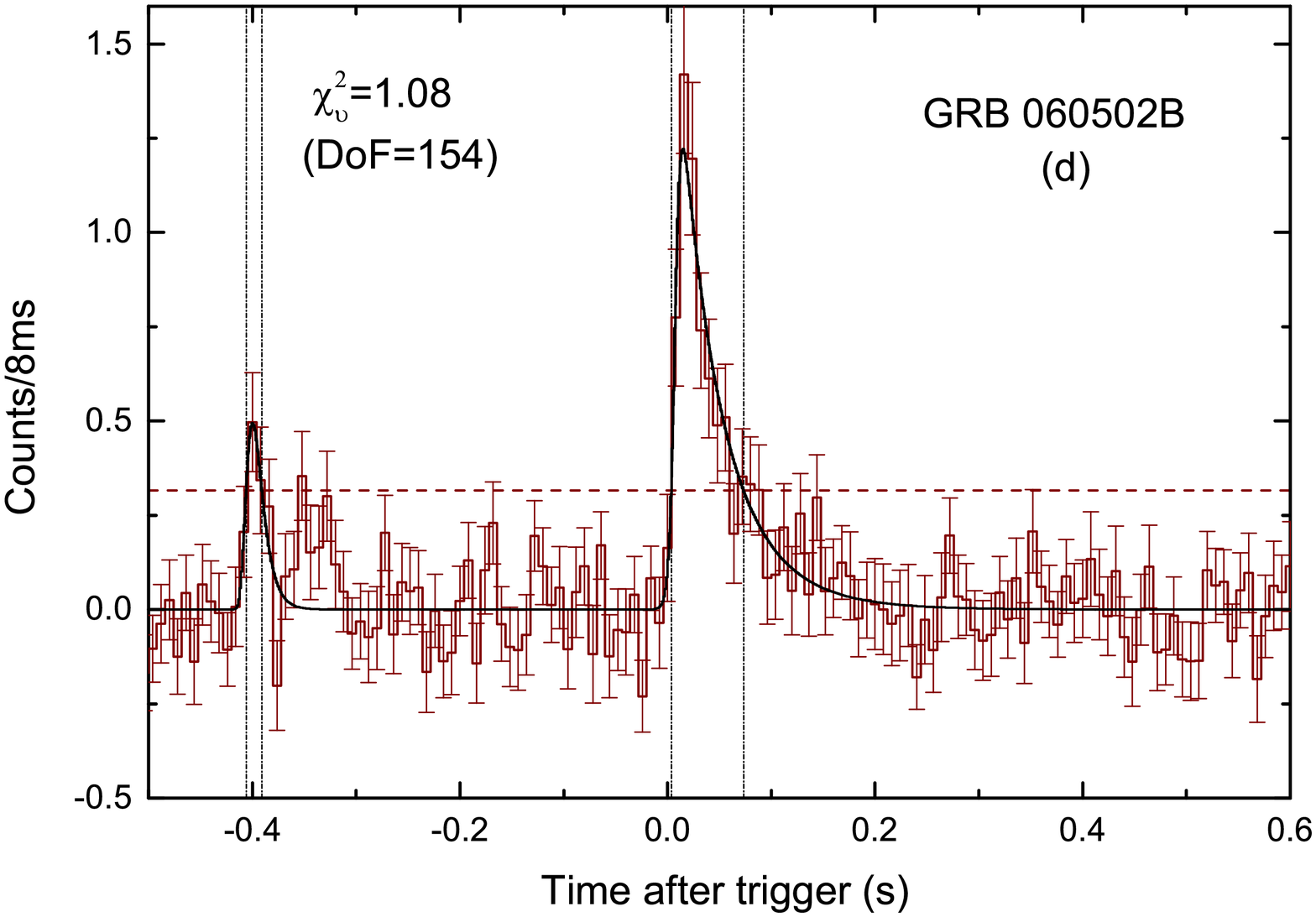}%"scale" 后的数字为图形的宽度，也可用"width=1.0\columnwidth" 定义
  \end{minipage}
    \begin{minipage}[c]{1\textwidth}
\includegraphics[height=4cm, width=7.8cm, angle=0]{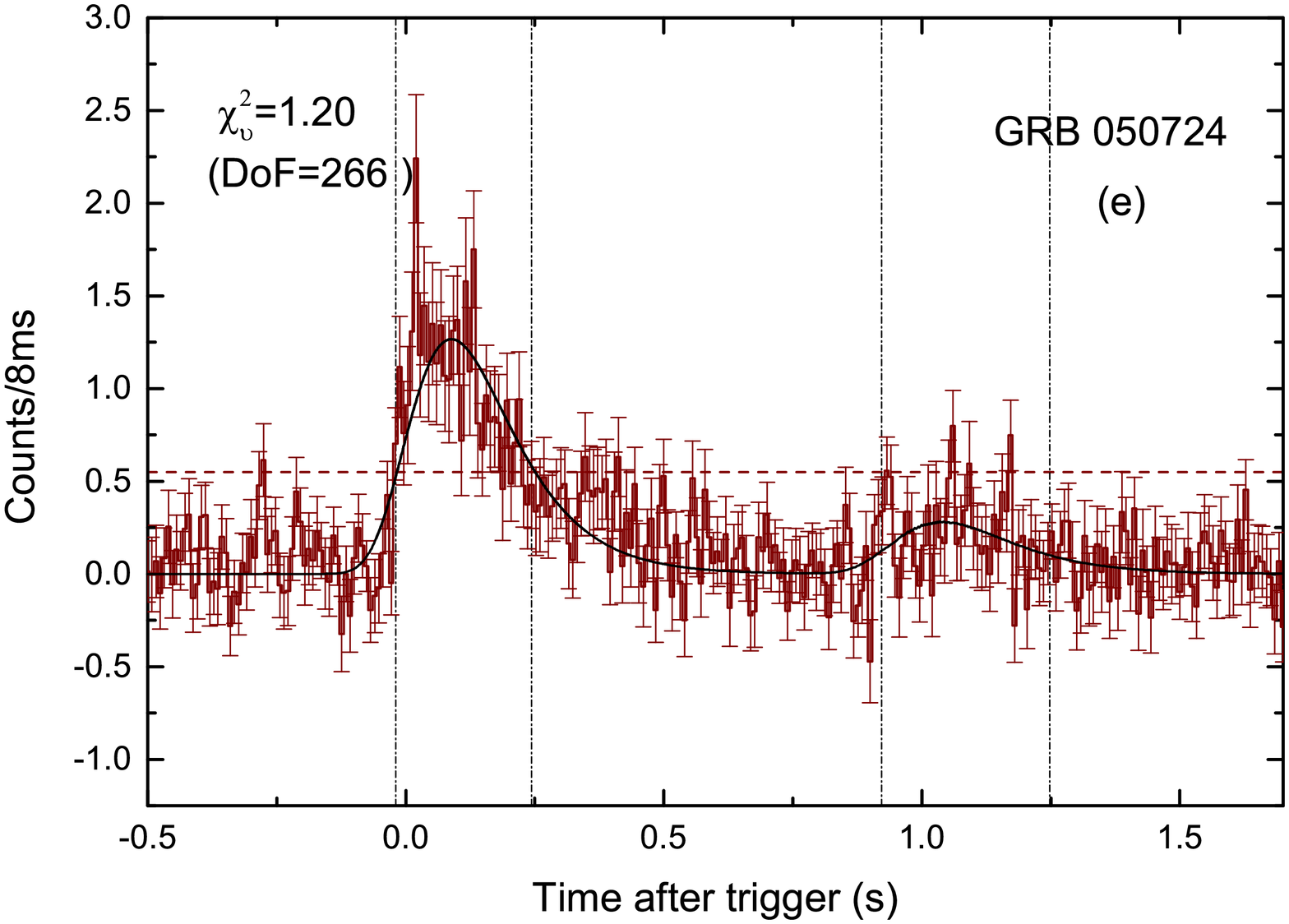}%"scale" 后的数字为图形的宽度，也可用"width=1.0\columnwidth"定义
  \end{minipage}
\caption{The light curves (15-350KeV) of five typical sGRBs. The horizontal
dashed lines mark a $3\sigma$ confidence level. The vertical dashed lines show the starting and the ending times of the emission target pulses. (a) single pulse; (b) M-loose; (c) M-tight; (d) Main peak with precursor; (e) Main peak with EE.} % 图题
\label{Fig1}%{}中"fig:example1"为图名，引用时用\ref{fig:example1}
\end{figure}

The mask-weighted light curve data of the sGRBs with a 8-ms resolution are taken from the Swift website \citep{Lien2016} \footnote{\url{https://swift.gsfc.nasa.gov/results/batgrbcat/}}. Note that all light curves of the selected 26 Swift/BAT sGRBs can be well fitted by the empirical Kocevski-Ryde-Liang (KRL) function that has been popularly used in literatures  \cite[e.g.,][]{Kocevski2003,ZhangandQin2005,Li2020,Li2021}. Fig.~\ref {Fig1} shows five representative cases that can be fitted successfully according to the reduced Chi-square standard together with a residual assessment \citep{Li2020,Li2021}. The KRL function with five free parameters is \begin{equation}\label{equation:1}
f(t)=f_m(\frac{t+t_0}{t_m+t_0})^r[\frac{d}{d+r}+\frac{r}{d+r}(\frac{t+t_0}{t_m+t_0})^{(r+1)}]^{-(\frac{r+d}{r+1})},
\end{equation}
where \emph{r} and \emph{d} respectively determine the rise and the decay shapes of an individual pulse, $f_m$ represents the peak flux, $t_m$ is the peak time, $t_0$ is the offset of the pulse from the trigger time. The fitting processes have been conducted in energy channel 15-350 keV.  We define the duration of an emission segment with \emph{t$_{dur}$=t$_{e}$-t$_{s}$}, in which \emph{t$_{s}$} and \emph{t$_{e}$} are the starting and the ending times of a given pulse at the level of S/N=3. Based on the above fitting with Eq. 1, one can easily  obtain all temporal features of different kinds of sGRBs. Finally, our sample includes 42 GRB pulses, of which 33, 7 and 2 pulses have been taken from main peaks, precursors and EEs, respectively. The fitting parameters are listed in Appendix A Table \ref{tab1}.

Subsequently, we utilize software \textsc{Xspec} to perform the model fitting of the spectrum of each episode as shown in Fig.~\ref {Fig1}. A power-law (PL) or cutoff power-law (CPL) spectral form has been applied to fit the GRB spectra because the \emph{Swift/BAT} has a narrow energy band \citep [see also][]{Zhangzb2020}. The PL model is written as
\begin{equation}\label{equation:2}
N_{E,PL}(E)=N_{0,PL} E^{-\alpha_{pl}},
\end{equation}
where \emph{N$_{0,PL}$ }is the photon flux (photons cm$^{-2}$ keV$^{-1}$ s$^{-1}$), \emph{E} is the photon energy, $\alpha_{pl}$ is the
low photon spectral index. The CPL model can be written as
\begin{equation}\label{equation:3}
N_{E,CPL}(E)=N_{0,CPL} E^{-\alpha_{cpl}} exp(-E/E_{p}),
\end{equation}
in which \emph{N$_{0,CPL}$} is the photon flux (photons cm$^{-2}$ keV$^{-1}$ s$^{-1}$), \emph{E} is the photon energy, $\alpha_{cpl}$ is the
low photon spectral index, \emph{E$_{p}$} is the peak energy in keV.
Besides, a Planck black-body (BB) function used to identify the thermal component can be
expressed by
\begin{equation}\label{4}
N_{E,BB}(E)=N_{0,BB}{\frac{8.0525E^2dE}{(kT)^4[exp(E/kT)-1]}},
\end{equation}
where $kT$ is the thermal energy of electrons and \emph{E} is the photon energy, both energies are in units of keV. Note that the model is built in Xspec.
%% Authors can give a citation as 'Michel et al. 1992'.

The reduced $\chi{^2_\nu}$ is given to estimate the goodness of spectrum fitting.
We choose CPL if $\Delta\chi{^2}$ = $\chi{^2_{PL}}-\chi{^2_{CPL}}>6$ and PL if $\chi{^2_{PL}}-\chi{^2_{CPL}}\leq6$ as the best-fit model \citep{Sakamoto2009,Lien2016,Katsukura2020}. This criterion is used in the BAT team for reporting the spectral parameters based on a CPL fit. Totally, Appendix A Table \ref{tab2} lists the results of the temporal and spectral properties of 26 typical Swift/BAT sGRBs. Column (1) lists the GRB name; Column (2) lists the duration $T_{90}$; Column (3) lists the cosmological redshift; Column (4) lists the duration \emph{t$_{dur}$} of each pulse; Columns (5) - (10) respectively represent the observed peak energy $E_{p}$, the spectral index $\alpha$, the average flux in unit of erg \ cm$^{-2}$ s$^{-1}$, the observed energy fluence in unit of erg cm$^{-2}$, and the goodness of spectrum fitting for the PL model; While columns (11) - (16) display the corresponding parameters for the CPL model. Finally, Columns (17) and (18) show the $\Delta\chi{^2}$ = $\chi{^2_{PL}}-\chi{^2_{CPL}}$ and the best model. Note that we use the time of the valleys as the boundary of overlapping pulses in Mt-DPs. Because of relatively weaker EE signal, the duration of sGRB 050724 is defined in terms of the time domain when the fitted intensity is equal to e$^{-1}$ of its maximum value. And note that our sample also includes the first gravitational-wave associated
GRB 170817A detected by Fermi/GBM \citep{Goldstein2017,Savchenko2017}. \cite{ZhangBB2018a} reported the detailed temporal and spectral properties for the main peaks and the EE components. Using their results, we compare the properties of GRB 170817A with those of the other typical sGRBs in our samples thus it is beneficial to identifying the candidates similar to GRB 170817A.
\section{Results}

\label{sect:Obs}
In this section, we present the main results of temporal and spectral
parameters such as pulse durations (\emph{t$_{dur}$}), peak energy ($E_{p}$), average flux ($F_{p}$), and energy fluence ($S_\gamma$) \footnote{The fluence is calculated in the energy range of 15-350 keV.} together with their correlations and evolutions.

%      One column rotated figure
%-------------------------------------------------------------

\subsection{Spectral characteristics of diverse prompt emission segments}
Most Swift/BAT GRB spectra can be fitted by a simple PL due to the narrow energy band \citep{Zhang2007a,Zhang2007b,Sakamoto2011}.  Previous studies illustrated that the relation $\alpha_{pl}$ $-$ \emph{E$_{p}$} can be employed as an indicator to estimate the \emph{E$_{p}$} of a burst without good spectral breaks \citep{Crider1997,Kaneko2006,Zhang2007b,Sakamoto2009,Virgili2012}.
This relation was first derived by \cite{Zhang2007b} and then confirmed by \cite{Virgili2012} as
\begin{equation}\label{equ5}
logE_{p}=(2.76\pm0.07)-(3.61\pm0.26) log\alpha_{pl} ,
\end{equation}
with a varied low energy spectral index of $1.2\leq\alpha_{pl}\leq2.3$. Similar to the conclusion of \cite{Sakamoto2009}, the bursts with $\alpha_{pl} <1.2 $ should have a higher $E_p$ far beyond the Swift/BAT band while the bursts with $\alpha_{pl} >2.3 $ are likely X-ray flashes with $E_p$ near or below the low-energy end of Swift/BAT \citep{Zhang2007a,Zhang2007b}.

In spite of the Eq. \ref{equ5} is heavily dominated by lGRBs, the sGRBs are generally consistent with the relationship \cite[see Fig. 2 in][]{Zhang2007b}.
Additionally, using a completely new GRB sample, including 31 short and 252 long GRBs with well-measured peak energy and redshift, \cite{ZhangBB2018a} found that short and long GRBs hold the coincident $E_{p,i} - E_{iso}$ correlations,
indicating that both kinds of GRBs may share the same radiation mechanism, which is consistent with the conclusion of \cite{Minaev2020}.
Consequently, we assume that Eq. \ref{equ5} is also available for sGRBs, together with their isolated emission components. Note that about 71\% (30/42) of the segmental spectra can be well fitted by the PL. There are 16 segments with  $1.2\leq\alpha_{pl}\leq2.3$, whose $E_p$ can be ideally fitted by the Eq. \ref{equ5}.

\subsubsection{GRB spectra in pulse durations}

\begin{figure}
\centering
\includegraphics[width=8cm, angle=0]{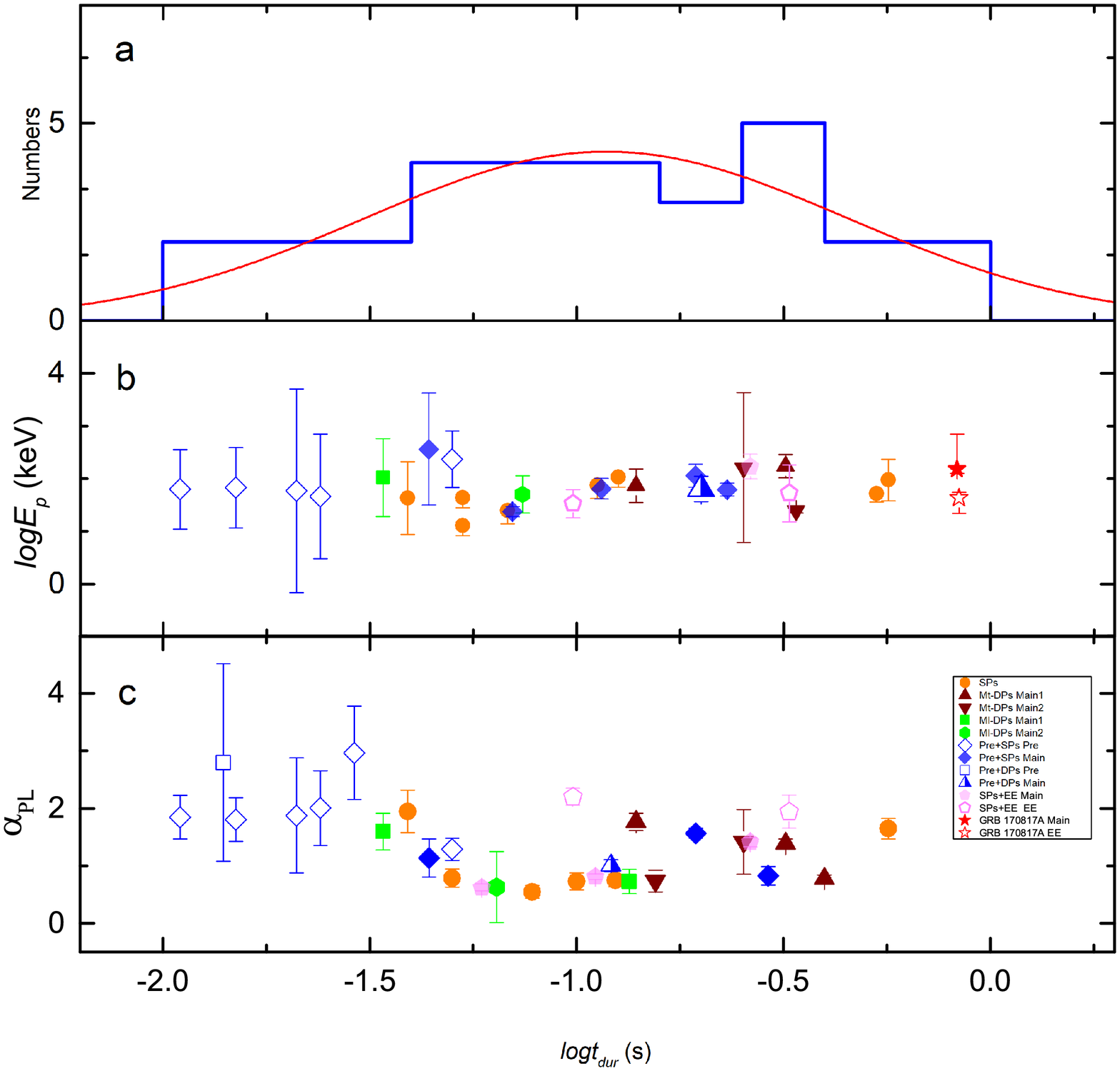}
   \caption{Panel a: distribution of t$_{dur}$ for 30 segments with $E_{p}$ measurement; Panel b: $E_{p}$ is plotted against \emph{t$_{dur}$} for 30 segments with $E_{p}$ measurement; Panel c: $\alpha_{PL}$ is plotted against \emph{t$_{dur}$} for 30 segments. The main peaks are marked with filled shapes. The precursors or the EEs are symbolized with empty shapes.}
   \label{tpeak}
   \end{figure}

We estimate the $E_p$ within each $t_{dur}$ of 30 isolated segments by way of either the Eq. 5 for the PL spectra or the CPL fitting directly. Fig. ~\ref{tpeak}a shows that the pulse durations are lognormally distributed with a mean value of $0.12\pm0.02$ s.
It is found that from Fig. ~\ref{tpeak}b \emph{$E_{p}$} and \emph{$t_{dur}$} are uncorrelated with each other due to a much lower correlation coefficient of 0.14. This is largely different from the anti-correlation of $E_{p}$ with $T_{90}$ between sGRBs and lGRBs \citep[e.g.,][]{zhangzb2008}. Moreover, we also analyze the relation between low energy spectral index $\alpha$ and pulse width in Fig. 2c, where no any correlation is found among them. No obviously spectral evolution across different prompt $\gamma$-ray components are manifested in Fig. 2b and 2c for the sGRBs. In general, the spectrum of late-time EEs is relatively softer than that of main peaks. However, almost all the EE segments identified in our sample occur within 2 seconds since the trigger time. It can be understood that the prompt sGRB spectra do not evolve in a very short period.

\subsubsection{Flux versus fluence}
\begin{figure}[H]%"[]"中为位置参数，四个参数tbph依次是置顶、置底、浮动、当前位置，，选用的参数优先顺序为h-t-b-p
\centering
  \includegraphics[width=8cm]{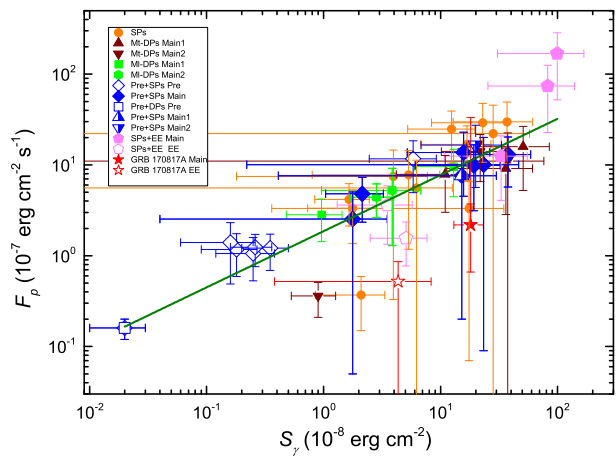}
\caption{$F_p$ is plotted against $S_\gamma$ for 44 pulses including GRB 170817A in our sample. The olive line denotes the best logarithmic fit. All symbols are same as in Fig. 2.} %图题
\label{fluxfluence}%{}中"fig:example1"为图名，引用时用\ref{fig:example1}
\end{figure}

Fig.~\ref{fluxfluence} shows a tight correlation between $F_p$ and $S_\gamma$.
The correlation for our selected sGRB sample is
\begin{equation}\label{equation:4}
log F_p=(0.62\pm0.07)log S_\gamma + (0.27\pm0.07),
\end{equation}
with a Pearson correlation coefficient $\rho=0.83$ and a chance probability $P=5.6\times10^{-12}$. It is necessary to announce that this correlation is valid only when all event pulses (including GRB 170817A) in our sample are considered.
Noticeably, the values of $F_p$ and $S_\gamma$ of the precursors are relatively lower than those of other components of sGRBs. It is worth nothing that GRB 170817A as an off-axis sGRB is marginally coincident with Eq. 6 and will affect this correlation slightly when it is ignored.
\subsubsection{Peak energy versus fluence}
\begin{figure}[H]
\centering
   \includegraphics[width=80mm]{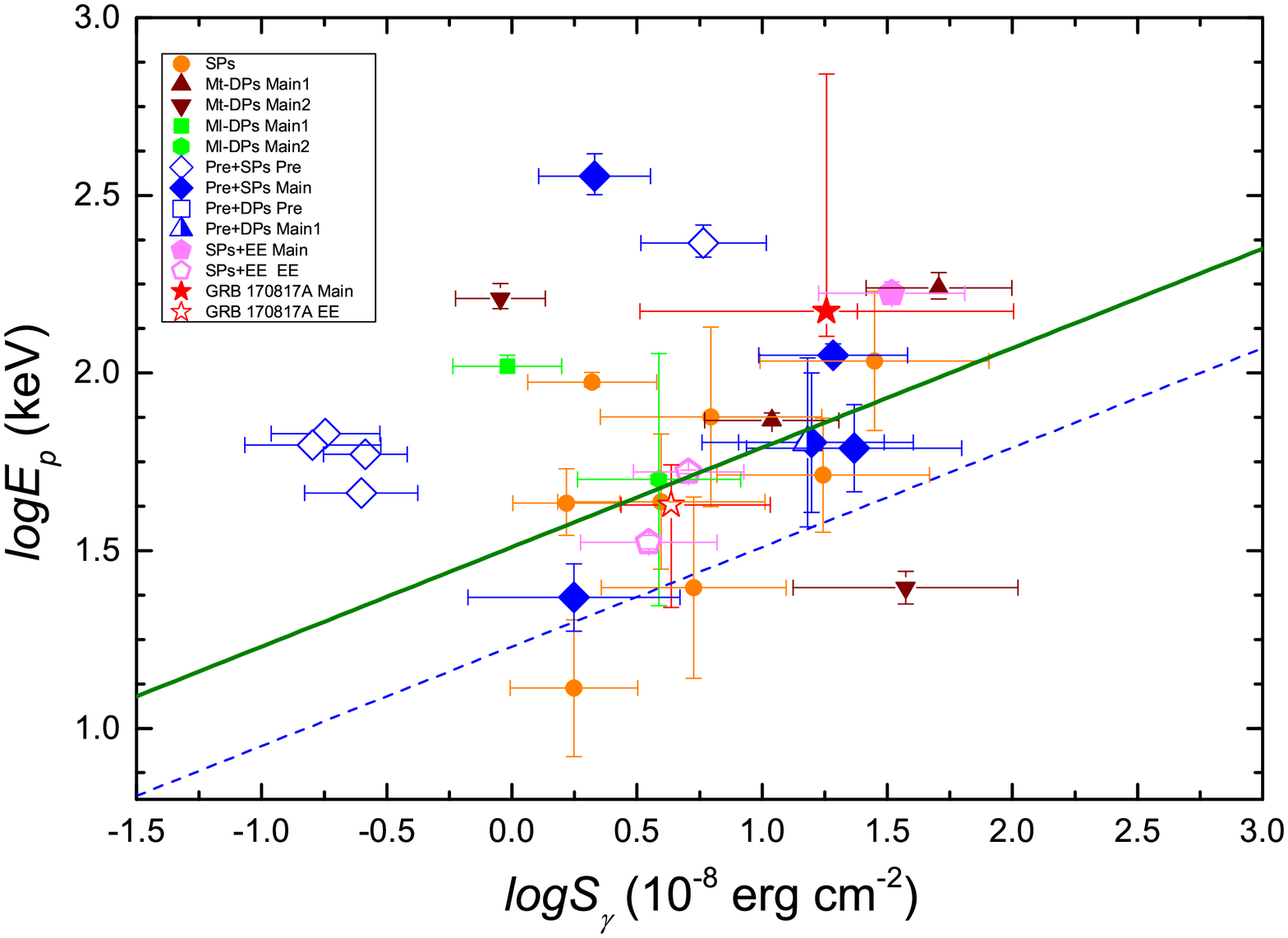}
     \caption{$E_{p}$ is plotted against $S_\gamma$ for 30 segments including GRB 170817A. The solid olive and dashed royal lines are the best logarithmic fit to the correlation between $S_\gamma$ and $E_p$ for the sGRBs and lGRBs in \cite{Zhangzb2020}. All symbols are same as in Fig. 2. }
   \label{fluencepeak}
\end{figure}

Recently, we studied the observed $E_{p}-S_\gamma$ relations of 283 Swift GRBs comprising 252 lGRBs and 31 sGRBs with known redshift and measured $E_{p}$ and found that sGRBs and lGRBs are differently distributed in the plane of $E_{p}$ versus S$_\gamma$ \citep{zhangzb2018}.
More recently, \cite{Zhangzb2020} proposed a useful correlation of $E_p \sim S_\gamma^{0.28}$ that can be applied as an $E_{p}$ indicator for those bursts with unknown $E_p$ \citep{Zhangzb2020}. Motivated by these results, we now focus on the analysis of the same correlation for the diverse emission segments in sGRBs whose light curves can be well fitted as shown in Fig. 1.  It is surprisingly shown in Fig.~\ref{fluencepeak} that there is no obvious correlation between $E_{p}$ and $S_\gamma$ with a Pearson coefficient of 0.16 and
a chance probability of 0.39, which is primarily resulted from the independence of peak energy on pulse duration as exhibited in Fig. 2. The solid and dashed lines show the empirical relation of $E_{p}$ and $S_\gamma$ of the sGRBs and lGRBs with well measured spectrum proposed by \cite{Zhangzb2020}. Note that GRB 170817A resides among the sGRB group, which is much similar to the finding for 31 sGRBs with known redshift by \cite{zhangzb2018} and the short or type E-II GRBs with EE in \cite{Zhangxiaolu2020}.


\subsection{Spectral evolution}
\subsubsection{Radiative intensity of diverse segments}

In order to test whether the spectra of sGRBs evolve from the early components to the later one in the phase of prompt $\gamma$-ray emissions, we compare the peak flux $F_{p}$ of the early pulse ($F_{p,e}$) with the later pulse ($F_{p,l}$), as well as other key spectral parameters ($E_p$ and $S_{\gamma}$).
\begin{figure}[H]
   \includegraphics[width=8cm, angle=0]{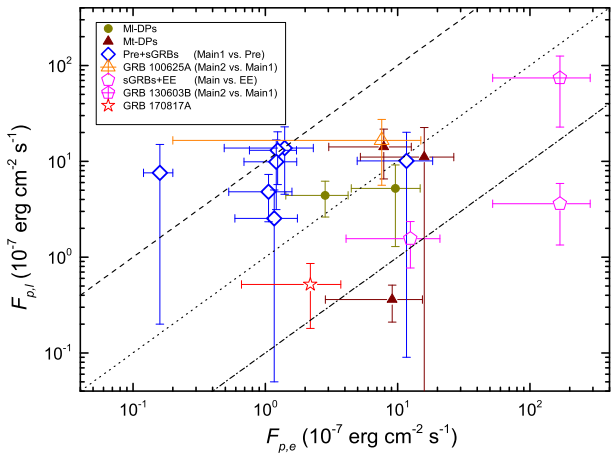}
   \centering
   \caption{$F_{p}$ comparisons between the early and the later pulses in the one-component (two-type DPs, filled symbols) and the two-component (empty symbols) sGRBs. Three peak flux ratios of $F_{p,l}$ to $F_{p,e}$ are signified by the dashed, dotted and dash-dotted lines for 10, 1 and 1/10, respectively. }
   \label{flux}
   \end{figure}
\begin{figure}[H]
   \includegraphics[width=8cm, angle=0]{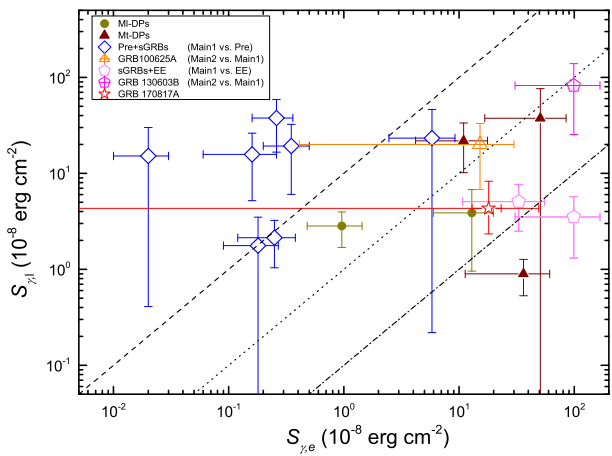}
   \centering
\caption{Comparison between $S_{\gamma}$ of the early and the later pulses in the one-component (two-type DPs, filled symbols) and the two-component (empty symbols) sGRBs. Three peak flux ratios of $S_{\gamma,l}$ to $S_{\gamma,e}$ are signified by the dashed, dotted and dash-dotted lines for 10, 1 and 1/10, respectively. }
   \label{fluence}
   \end{figure}

We compare the peak flux densities of main peaks (filled symbols) with both precursors and  EEs (empty symbols) in Fig.~\ref{flux}, where it can seen that the main peaks are on average brighter than the other two components about one order of
magnitude \citep[see also][]{Zhangxiaolu2020}. In addition, two types of DPs lying near to the dotted line, shows that the brightness of two main peaks are comparable with each other, which is consistent with the conclusion drawn by \cite{Lan2018}. Based on an analysis of Pre+Mt-DP 100625A and Mt-DP+EE 130603B, we find that the peak fluxes of two main peaks are similar to those isolated Mt-DPs. Furthermore, we find the similar results for the observed fluence $S_{\gamma}$ in Fig. ~\ref{fluence}.
%\begin{figure}[H]
%   \centering
%   \includegraphics[width=8.5cm, angle=0]{averageep.eps}
%   \caption{The $E_p$ evolution of different emission segments from the early to the late epochs. The black squares and red pentagons represent the mean and median values for 7 precursors, 12 SPs, 15 Main1 peaks, 7 Main2 peaks, and 3 EEs, respectively. }
%   \label{average}
 %\  \end{figure}
\textbf{\subsubsection{Features of spectral evolution}}

\cite{Zhang2007a} reported that the burst with $\alpha_{pl}>2.3$ is likely a softer $\gamma$ event called
XRF with peak energy near or below the low-energy end of BAT. In this case, Eq. \ref{equ5} cannot be used to estimate the $E_p$. For two precursors with $\alpha_{pl}>2.3$ of GRBs 100702A and 100625A, we try to invoke
the Planck black-body model (Eq. \ref {4}) to fit the spectra and obtain the thermal energy of electrons $KT=6.02\pm1.57 KeV$ and $KT=9.04\pm1.98 KeV$, indicating that the thermal contributions to GRB spectra are negligible. On the other hand, it is found that there are 12 main peaks whose best models are PL with $\alpha_{pl}<1.2$, including SPs, Pre+SPs, Pre+DPs and sGRB+EE. In this situation, Eq. \ref{equ5} cannot be used to also estimate the $E_p$ effectively, since the $E_p$ will be quite outside of the BAT band. Moreover, we find that the EEs of GRBs 050724 and 130603B are slightly softer than their main peaks, which is consistent with some previous conclusions \citep[e.g.,][]{Norris2006,Norris2010,Kagawa2015}.
However, we emphasize that the spectra of sGRBs in our sample do not evolve during prompt $\gamma$-ray emission epoch, which challenges the known theoretical models for the precursors and the EEs \citep [e.g.][]{Murakami1991, Lyutikov2000, Metzger2011}. On the other hand, the non-evolutionary phenomena can be supported by the zero lags of light curves between different energy channels for sGRBs (e.g. Norris et al. 2006; Zhang et al. 2006). By looking back to Fig. 2, we can conclude that the parameters $E_p$ and $\alpha$ are two representative qualities describing the spectral evolution consistently \citep [see also][]{Ghirlanda2004}.
\section{Conclusions and Discussions}
We summarize the major results as follows.

(1) We find that the peak energies of diverse $\gamma$-ray radiation segments in sGRBs with single or double main peaks are uncorrelated with the corresponding $t_{dur}$.

(2) We find a tight correlation between $F_p$ and $S_\gamma$ for different segments to be $log F_p=(0.62\pm0.07) log S_\gamma + (0.27\pm0.07)$.

(3) In the plane of $E_p$ versus $S_\gamma$, these diverse $\gamma$-ray radiation segments in sGRBs distribute near to the line of $E_p\sim S_\gamma^{0.28}$ found by \cite{Zhangzb2020} for those sGRBs with a well-measured spectrum. However, there is no obvious correlation found between $E_p$ and $S_\gamma$ for these segments entirely.

(4) The main peaks are on average brighter than the precursors or EEs about one order of magnitude. Regarding the EEs, our result is consistent with \cite{Zhangxiaolu2020}.

(5) In terms of the analyses of peak energies and low spectral index of diverse $\gamma$-ray radiation segments, it is found that the sGRB spectra of precursors, main peaks and EEs exhibit no obvious evolutionary sequence.

Unfortunately, since the absences of the EEs or precursors might be related to sensitivity or energy coverage of the current GRB detectors, no such sGRBs with three distinct
components have been observed. For example, though the Fermi/GBM with a broader energy band had identified over 2000 GRBs \citep{Kienlin2020}, only 4 of 244 precursors are identified in sGRBs \citep{Coppin2020}. Fortunately, more and more GRB monitors have been launched or planned to launch to meet the increasing requirements of the X-ray/gamma-ray counterpart observation.
More than 200 GRBs were detected by Chinese first X-ray astronomical satellite Hard X-ray Modulation Telescope (HXMT), thanks to its wider energy coverage from 1 keV to 3 MeV, large filed of view, and good sensitivity \citep{zhangshuangnan2020,liu2020}. The Gravitational wave high-energy Electromagnetic Counterpart
All-sky Monitor (GECAM) which has an all-sky field of view, a high sensitivity and a wide energy interval (6 keV - 5 MeV) has been launched in 2020 \citep{Liao2020,Song2020,Chen2020}. Meanwhile, Space multi-band astronomical Variable Objects Monitor (SVOM) whose energy range is from 15 keV to 5 MeV aims at detecting very distant and faint/soft nearby GRBs. SVOM with rapid slew capability will provide GRB positions and spectral parameters on very short time scale in the near decade through a collection of instruments in various gamma and X energy bands as well as in visible wave lengths through a narrow field of view telescope \citep{Wei2016}.
Hopefully, our results can shed new light on the studies of physical processes of sGRBs. Meanwhile, further search for three-components sGRBs simultaneously from the Fermi, HXMT, GECAM and SVOM catalogs,
can draw more robust conclusions in the future.

\begin{acknowledgements}
We thank the anonymous referee for very helpful suggestion and
constructive comments. This work was supported by the National Natural Science Foundation of China (No. U2031118), the Youth Innovations and Talents Project of Shandong Provincial Colleges and
Universities (Grant No. 201909118) and the Natural Science Foundations (ZR2018MA030, XKJJC201901).
\end{acknowledgements}
%\tabcolsep 2pt


%\rotatebox{90}{Appendix A:Tables}
\centering
Appendix A:Tables
\begin{sidewaystable*}[!h]
\tiny
\centering
\setlength{\abovecaptionskip}{0pt}
\setlength{\belowcaptionskip}{-5pt}
\caption{Fitting parameters of all individual pulses in Swift sGRBs.}
\tabcolsep 2pt
\begin{tabular}{llllllll}
\specialrule{0em}{10pt}{1pt}
\hline
\hline
GRB  & $f_m$  &$t_m$ & r& d& $t_0$ & $DOF$ & $\chi^2/DOF$\\
\hline
\hline
SPs&&&&&&&\\\hline
070923&$2.053\pm0.483$ & $0.005\pm0.004$ & $155.833\pm177.285$ & $13.529\pm2.156$ &$0.300\pm0.000$& 84 &1.36\\\hline
090621B&$1.106\pm0.139$ & $0.053\pm0.006$ & $70.672\pm1055.989$ & $34.488\pm459.488$ &$1.127\pm16.704$& 97 &1.13\\\hline
100206A&$1.785\pm0.144$ & $0.039\pm0.006$ & $22.181\pm118.060$ & $10.749\pm39.672$ &$0.400\pm2.004$& 90 &0.90\\\hline
110420B&$2.601\pm3.424$ & $0.002\pm0.020$ & $(-2.627\pm91.410)E14$ & $(-3.722\pm12.307)E12$ &$(-1.209\pm4.090)E11$& 73 &1.21\\\hline
120305A&$5.831\pm0.294$ & $0.020\pm0.001$ & $63.766\pm7.074$ & $11.159\pm0.629$ &$0.300\pm0.000$& 85 &
1.23\\\hline
131004A&$0.661\pm0.035$ & $0.051\pm0.018$ & $3.537\pm0.515$ & $2.748\pm0.372$ &$0.500\pm0.000$& 248 &1.11\\\hline
140622A&$0.717\pm0.126$ & $0.016\pm0.009$ & $14.893\pm79.699$ & $3.965\pm10.719$ &$0.151\pm0.678$& 176 &0.96\\\hline
150301A&$3.085\pm0.237$ & $0.006\pm0.001$ & $175.122\pm46.756$ & $15.547\pm1.640$ &$0.290\pm0.000$& 59 &0.94\\\hline
150710&$2.356\pm1.744$ & $0.016\pm0.016$ & $(3.732\pm18.496)E14$ & $(6.923\pm6.081)E12$ &$(2.607\pm2.116)E11$& 71 &1.19\\\hline
160601A&$0.875\pm0.159$ & $0.056\pm0.008$ & $4.582\pm3.901$ & $(-1.211\pm3.133)E13$ &$0.181\pm0.165$& 96 &0.98\\\hline
180727A&$0.587\pm0.036$ & $0.505\pm0.018$ & $3.769\pm0.227$ & $(-1.497\pm1.246)E13$ &$0.900\pm0.000$& 434 &1.07\\\hline
190326A&$4.922\pm0.548$ & $0.006\pm0.002$ & $263.792\pm1148.656$ & $15.800\pm58.584$ &$0.409\pm1.669$& 90 &1.00\\\hline \hline
Mt-DPs&&&&&&&\\\hline
101219A&$1.527\pm0.092$ & $0.020\pm0.007$ & $63.213\pm48.603$ & $3.882\pm1.583$ &$0.728\pm0.437$& 367 &0.93\\ \hline
       &$0.551\pm0.145$ & $0.541\pm0.009$ & $19.311\pm2.346$ & $(-8.259\pm1.398)E14$ &$0.761\pm0.158$&-- &--\\ \hline
120804A&$1.447\pm0.129$ & $0.106\pm0.010$ & $-5.132\pm0.528$ & $(-8.984\pm1.414)E13$ &$0.999\pm0.132$&379 &1.06\\ \hline
&$1.718\pm0.214$ & $0.402\pm0.007$ & $25.034\pm11.678$ & $(-2.131\pm2.000)E14$ &$1.002\pm0.622$&  -- &--\\ \hline
130912A&$1.952\pm0.501$ & $0.033\pm0.003$ & $-22.117\pm30.011$ & $(-1.537\pm1.981E14$ &$0.487\pm0.686$& 115 &1.19\\ \hline
&$0.794\pm0.390$ & $0.271\pm0.008$ & $18.102\pm10.586$ & $(-1.448\pm1.966)E14$ &$0.631\pm0.494$&-- &--\\ \hline \hline
Ml-DPs &&&&&&&\\\hline
 120229A&$0.650\pm0.150$ & $0.043\pm0.006$ & $79.069\pm284.368$ & $(-1.130\pm3.963)E16$ &$0.986\pm3.571$&116&1.00\\ \hline
 &$0.421\pm0.081$ & $0.215\pm0.006$ & $25.489\pm2.742$ & $(-1.851\pm0.206)E15$ &$0.772\pm0.101$&  -- &--\\ \hline
111117A&$0.439\pm0.099$ & $0.107\pm0.014$ & $-8.705\pm4.986$ & $(-1.139\pm0.717)E14$ &$0.598\pm0.429$& 155 &1.21\\ \hline
&$0.740\pm0.311$ & $0.487\pm0.007$ & $27.929\pm13.745$ & $(2.971\pm7.309)E14$ &$0.254\pm0.302$&  -- &--\\ \hline\hline
%Two-component sGRBs&&&&&&&&&&&&&&&&\\ \hline\hline
Pre+SPs&&&&&&&\\ \hline
060502B&$0.494\pm0.165$ & $-0.340\pm0.004$ & $-38.813\pm33.342$ & $(1.317\pm1.080)E15$ &$0.700\pm0.182$& 154 &1.08\\ \hline
&$1.222\pm0.154$ & $0.015\pm0.003$ & $254.514\pm1745.588$ & $18.119\pm108.761$ &$0.700\pm4.563$&  -- &--\\ \hline
071112B&$0.438\pm1.325$ & $-0.568\pm0.005$ & $20.961\pm25.776$ & $(-1.424\pm1.410)E15$ &$0.936\pm0.427$& 204 &1.04\\ \hline
&$0.403\pm0.067$ & $0.075\pm0.007$ & $26.953\pm8.395$ & $(-6.233\pm3.660)E14$ &$1.012\pm0.305$&  -- &--\\ \hline
100702A&$0.548\pm0.231$ & $-0.250\pm0.004$ & $19.330\pm17.660$ & $(-7.487\pm7.021)E14$ &$0.521\pm0.242$& 116 &0.93\\ \hline
&$1.482\pm0.101$ & $0.086\pm0.003$ & $-26.920\pm13.772$ & $(1.872\pm1.110)E14$ &$1.241\pm0.728$&  -- &--\\ \hline
160408A&$0.552\pm1.278$ & $-0.928\pm0.105$ & $79.970\pm956.924$ & $1.692\pm2.753$ &$0.997\pm0.235$&316 &1.06\\ \hline
&$0.646\pm0.106$ & $0.242\pm0.016$ & $94.775\pm22.713$ & $(-1.616\pm0.743)E15$ &$12.686\pm3.267$&  -- &--\\ \hline
160726A&$1.474\pm0.564$ & $0.020\pm0.003$ & $-68.274\pm10.965$ & $(-4.360\pm1.310)E15$ &$0.998\pm0.171$&  242 &1.14\\ \hline
&$1.616\pm0.151$ & $0.617\pm0.005$ & $-24.922\pm21.999$ & $(-7.269\pm11.494)E14$ &$1.000\pm1.487$&  -- &--\\ \hline
180402A&$1.359\pm65.982$ & $-0.203\pm0.514$ & $463.348\pm61482.680$ & $25.266\pm1759.530$ &$0.488\pm21.032$& 129 &1.12\\ \hline
&$0.923\pm0.378$ & $0.195\pm0.009$ & $9.312\pm1.557$ & $(-7.793\pm1.723)E14$ &$0.498\pm0.106$&  -- &--\\ \hline\hline
Pre+Mt-DPs&&&&&&&\\\hline
100625A&$0.448\pm3.709$ & $-0.376\pm0.040$ & $200.145\pm13076.606$ & $(1.017\pm105.800)E15$ &$1.453\pm65.833$& 205 &1.38\\ \hline
&$0.777\pm0.121$ & $0.048\pm0.008$ & $-201.869\pm199.456$ & $(-1.336\pm2.153)E15$ &$-10.802\pm10.544 $& -- &--\\ \hline
&$1.058\pm0.086$ & $0.213\pm0.009$ & $(3.844\pm6.463)E6$ & $(-6.166\pm15.290)E18$ &$(2.648\pm4.395)E5$&  -- &--\\ \hline\hline
sGRBs+EE&&&&&&&\\ \hline
050724&$1.256\pm0.123$ & $0.084\pm0.006$ & $(2.524\pm0.409)E13$ & $(-1.043\pm0.809)E26$ &$(-2.498\pm0.391)E12 $&266 &1.20\\ \hline
&$0.280\pm0.055$ & $1.043\pm0.025$ & $338.931\pm37.597$ & $(7.507\pm0.424)E16$ &$-40.226\pm4.399$& -- &--\\\hline
130603B&$15.022\pm2.499$ & $0.017\pm0.004$ & $86.579\pm129.604$ & $5.348\pm4.760$ &$0.112\pm0.148$& 174 &1.17\\ \hline
&$4.828\pm0.832$ & $0.072\pm0.002$ & $16.574\pm226.295$ & $14.620\pm140.411$ &$0.108\pm2.183 $& -- &--\\ \hline
&$0.574\pm0.152$ & $0.192\pm0.010$ & $8.537\pm0.765$ & $(9.342\pm0.700)E14$ &$0.384\pm0.041$&  -- &--\\ \hline
\hline
\end{tabular}
\label{tab1}
\end{sidewaystable*}

\begin{sidewaystable*}[h]
\tiny
\centering
\setlength{\abovecaptionskip}{0pt}
\setlength{\belowcaptionskip}{-5pt}
\caption{Characteristic Parameters of Diverse Gamma-ray Segments in sGRBs}
\tabcolsep 2pt
\begin{tabular}{lccccccccccccccccccc}
\specialrule{0em}{10pt}{1pt}
\hline
$GRB$&\emph{$T_{90}$}&redshift&\emph{$t_{dur}$} & $Model$& $E_{p}$$^a$& $Index$&$Flux$ $^d$& $Fluence^e$&$\chi^2/DOF$& $Model$&$E_{p}$& $Index$  &  $Flux$ $^d$& $Fluence^e$&$\chi^2/DOF$&$\bigtriangleup\chi^2$&best-fit  \\
  & s&z&$(t_s,t_e)$&&\emph{keV}&$\alpha$ & &&&&\emph{keV}&$\alpha$& &&&&Model\\
(1)&(2)&(3)&(4)&(5)&(6)&(7)&(8)&(9)&(10)&(11)&(12)&(13)&(14)&(15)&(16)&(17)&(18)\\
\hline
\hline
SPs&&&&&&&&&&&&&&&\\\hline
070923&0.040&-&(-0.003,0.047)&PL&&\textbf{0.79$\pm$0.16$^c$}&24.90$\pm$14.42&12.40$\pm$7.19&66.02/73 &CPL&100.11$\pm$100.39 &0.17$\pm$0.63 &-&-&64.54/72 &1&PL\\\hline
090621B&0.140&-&(0.011,0.111)&PL&&\textbf{0.73$\pm$0.15$^c$} &16.82$\pm$10.08&16.85$\pm$10.10&68.39/73 &CPL&135.29$\pm$156.51 &0.26$\pm$0.56 &-&-&67.08/72 &1&PL\\\hline
100206A&0.116&0.4068&(-0.009,0.115)&PL&&\textbf{0.75$\pm$0.11$^c$} &29.85$\pm$19.52&37.02$\pm$24.20&72.02/73 &CPL&232.22$\pm$357.46 &0.49$\pm$0.40 &-&-&71.17/72 &1&PL\\\hline
110420B&0.084&-&(-0.000,0.052)&PL&-&2.39$\pm$0.36 &-&-&121.77/73 &CPL&12.98$\pm$5.76 &-1.70$\pm$1.50 &3.34$\pm$1.97&1.77$\pm$1.04&99.41 /72 &22&CPL\\\hline
120305A&0.100&-&(-0.005,0.122)&PL&-&1.13$\pm$0.06 &-&-&63.73/73 &CPL&108.05$\pm$48.70 &0.61$\pm$0.24 &22.21$\pm$23.39&28.09$\pm$29.58&57.43 /72 &($>$) 6&CPL\\\hline
131004A&1.536&0.717&(-0.144,0.386)&PL&-&1.77$\pm$0.07&-&-&87.72/73 &CPL&51.57$\pm$19.01 &0.84$\pm$0.32 &3.31$\pm$3.24&17.54$\pm$17.18&77.77 /72 &10&CPL\\\hline
140622A&0.13&0.959&(0.000,0.040)&PL&\emph{51.17$\pm2.71$}&\emph{1.95$\pm$0.37}&4.18$\pm$2.06&1.65$\pm$0.81&46.13/73 &CPL&79.36$\pm$227.53 &1.42$\pm$1.59 &-&-&45.96 /72 &0&PL\\\hline
150301A&0.484&-&(-0.002,0.051)&PL&-&1.70 $\pm$0.11 &-&-&77.69/73 &CPL&43.43$\pm$19.02 &0.55$\pm$0.49 &7.47$\pm$7.14&3.95$\pm$3.77&68.66 /72 &9&CPL\\\hline
150710&0.152&-&(0.012,0.090)&PL&&\textbf{0.55$\pm$0.11$^c$} &29.20$\pm$18.51&22.86$\pm$14.49&87.41 /73 &CPL&95.12$\pm$61.42 &-0.16$\pm$0.46 &-&-&83.77/72 &4&PL\\\hline
160601A&0.120&-&(-0.006,0.106)&PL&-&1.47$\pm$0.13 &-&-&90.92/73 &CPL&75.20$\pm$43.70 &0.54$\pm$0.42 &5.58$\pm$5.68 &6.24$\pm$6.35&78.43 /72 &12&CPL\\\hline
180727A&1.056&-&(0.205,0.771)&PL&\emph{94.23$\pm1.78$}&\emph{1.65$\pm$0.18} &0.37$\pm$0.22&2.09$\pm$1.24&77.92/73 &CPL&32.87$\pm$21.57 &0.22$\pm$0.88 &-&-&72.39 /72 &($<$) 6&PL\\\hline
190326A&0.076&-&(-0.002,0.067)&PL&-&2.66$\pm$0.19 &-&-&97.22 /73 &CPL&24.85$\pm$14.61 &0.85$\pm$0.93 &7.79$\pm$6.61&5.32$\pm$4.52&86.02 /72 &11&CPL\\\hline \hline
Mt-DPs &&&&&&&&&&&&&&&\\\hline
101219A&0.828&0.718&(-0.028,0.369)&PL&&\textbf{0.78$\pm$0.06$^c$} &9.12$\pm$6.27&36.22$\pm$24.90&63.97/73 &CPL&500.00$\pm$1559.14 &0.65 $\pm$0.21&-&-&64.23/72 &0&PL\\ \hline
              &&&(0.369,0.622)&PL&\emph{162.29$\pm7.26$}&\emph{1.42$\pm$0.56} &0.36$\pm$0.15&0.90$\pm$0.37&55.82/73 &CPL&213.82$\pm$1999.98 &1.15$\pm$2.09 &-&-&55.81/72 &0&PL\\ \hline
120804A&0.808&1.3&(-0.126,0.194)&PL&\emph{173.51$\pm1.15$}&\emph{1.39$\pm$0.08} &15.90$\pm$10.66&50.82$\pm$34.08&91.45/73 &CPL&133.82$\pm$93.43 &0.95$\pm$0.32 &-&-&89.15/72 &2&PL\\ \hline
              &&&(0.194,0.532)&PL&-&2.31$\pm$0.06 &-&-&269.15/73 &CPL&24.87$\pm$2.63 &-0.70$\pm$0.26 &11.06$\pm$11.44&
              37.45$\pm$38.74&81.86/72 &187&CPL\\ \hline
130912A&0.284&-&(0.005,0.144)&PL&\emph{73.51$\pm1.38$}&\emph{1.77$\pm$0.15} &7.87$\pm$4.85&10.94$\pm$6.75&71.36/73 &CPL&72.25$\pm$53.86 &0.94$\pm$0.54 &-&-&66.98/72 &4&PL\\ \hline
              &&&(0.144,0.299)&PL&&\textbf{0.74$\pm$0.19$^c$} &14.10$\pm$7.54&21.84$\pm$11.68&60.81/73&CPL&487.09$\pm$4493.18 &0.59$\pm$0.69 &-&-&60.64/72 &0&PL\\ \hline \hline
Ml-DPs&&&&&&&&&&&&&&&\\\hline
 120229A&0.22&-&(0.023,0.057)&PL&\emph{104.58$\pm3.43$}&\emph{1.60$\pm$0.32} &2.83$\pm$1.40&0.96$\pm$0.48&80.75/73 &CPL&111.33$\pm$225.58 &1.01$\pm$1.17 &-&-&80.38/72 &0&PL\\ \hline
              &&&(0.178,0.242)&PL&&\textbf{0.63$\pm$0.62$^c$} &4.41$\pm$1.79&2.83$\pm$1.14&73.54/73 &CPL&27.17$\pm$49.45 &-1.26$\pm$3.48 &-&-&71.73/72 &2&PL\\ \hline
111117A&0.464&2.211&(0.051,0.185)&PL&&\textbf{0.73$\pm$0.21$^c$} &9.64$\pm$5.18&12.92$\pm$6.94&78.87/73 &CPL&139.07$\pm$313.14 &0.31$\pm$0.76 &-&-&77.09/72 &2&PL\\ \hline
              &&&(0.442,0.516)&PL&-&1.23$\pm$0.20 &-&-&79.12/73 &CPL&50.13$\pm$41.00 &0.00$\pm$0.79 &5.23$\pm$3.94 &3.87$\pm$2.91&69.60/72 &10&CPL\\ \hline
%Two-component sGRBs&&&&&&&&&&&&&&&&\\ \hline\hline
Pre+SPs&&&&&&&&&&&&&&&&\\ \hline
060502B&0.144&0.287&Pre:(-0.406,-0.391)&PL&\emph{67.50$\pm3.22$}&\emph{1.81$\pm$0.38} &1.17$\pm$0.58&0.18$\pm$0.09&67.83/73 &CPL&495.33$\pm$6729.11 &1.71$\pm$1.42 &-&-&67.98/72 &0&PL\\ \hline
&&&Main:(0.004,0.073)&PL&-&2.27$\pm$0.15 &-&-&110.66/73 &CPL&23.34$\pm$5.10&-0.81$\pm$0.60 &2.54$\pm$2.49&1.77$\pm$1.73
&70.09/72 &41&CPL\\ \hline
071112B& 0.304&-&Pre:(-0.581,-0.557)&PL&\emph{45.96$\pm4.52$}&\emph{2.01$\pm$0.65} &1.06$\pm$0.53&0.25$\pm$0.13&71.92/73 &CPL&16.28$\pm$15.35 &-1.49$\pm$3.11 &-&-&67.30 /72 &5&PL\\ \hline
 & &&Main:(0.050,0.094)&PL&\emph{358.43$\pm6.25$}&\emph{1.14$\pm$0.33}&4.82$\pm$2.48&2.14$\pm$1.10&69.56/73 &CPL&38.59$\pm$42.46 &-0.54$\pm$1.57 &-&-&64.56/72 &5&PL\\ \hline
 100702A&0.512&-&Pre:(-0.267,-0.238)&PL&&\textbf{2.97$\pm$0.81$^b$} &1.21$\pm$0.52&0.35$\pm$0.15&59.88/73 &CPL&5.94$\pm$9.16 &-1.97$\pm$7.33 &-&-&58.22/72 &2&PL\\ \hline
&&&Main:(0.020,0.214)&PL&\emph{112.22$\pm1.05$}&\emph{1.57$\pm$0.09} &9.91$\pm$6.78&19.23$\pm$13.17&75.79/73 &CPL&66.97$\pm$32.86 &0.76$\pm$0.40 &-&-&71.07 /72 &5&PL\\ \hline
160408A&0.320&-&Pre:(-0.931,-0.910)&PL&\emph{59.09$\pm7.89$}&\emph{1.88$\pm$1.00} &1.24$\pm$0.48&0.26$\pm$0.10&27.96/73 &CPL&10.13$\pm$69.94&1.22$\pm$14.00 &-&-&28.84/72 &1&PL\\ \hline
&&&Main:(0.073,0.362)&PL&
&\textbf{0.83$\pm$0.16$^c$} &13.01$\pm$7.28&37.71$\pm$21.10&52.75/73 &CPL&500.00$\pm$4036.87 &0.70$\pm$0.57 &-&-&52.99/72 &0&PL\\ \hline
160726A&0.728&-&Pre:(0.001,0.051)&PL&\emph{231.93$\pm2.93$}&\emph{1.29$\pm$0.19} &11.70$\pm$6.75&5.83$\pm$3.37&66.58/73 &CPL&54.8$\pm$41.53 &0.14$\pm$0.80&-&-&62.61 /72 &4&PL\\ \hline
&&&Main:(0.535,0.766)&PL&-&1.37$\pm$0.08 &-&-&84.00/73 &CPL&61.39$\pm$17.32 &0.23$\pm$0.30 &10.08$\pm$9.99&23.30$\pm$23.08&61.75 /72 &22&CPL\\ \hline
180402A&0.180&-&Pre:(-0.205,-0.194)&PL&\emph{62.71$\pm3.12$}&\emph{1.85$\pm$0.38} &1.40$\pm$0.91&0.16$\pm$0.10&78.22/73 &CPL&72.98$\pm$130.89 &1.09$\pm$1.40 &-&-&77.36 /72 &1&PL\\ \hline
&&& Main:(0.135,0.250)&PL&-&0.47$\pm$0.22 &-&-&147.13/73 &CPL&81.52$\pm$81.74 &-0.41$\pm$0.91 &13.72$\pm$9.19&15.73$\pm$10.54&68.46/72 &79&CPL\\ \hline\hline
Pre+Mt-DPs&&&&&&&&&&&&&&&\\ \hline
100625A&0.332&0.452&Pre:(-0.384,-0.370)&PL&
&\textbf{2.80$\pm$1.72$^b$} &0.16$\pm$0.04&0.02$\pm$0.01&37.91/73 &CPL&14.63$\pm$58.38 &0.00$\pm$9.95 &&&36.98 /72 &1&PL\\\hline
&&&Main1:(-0.059,0.141)&PL&-&0.99$\pm$0.09 &-&-&87.56/73 &CPL&63.73$\pm$39.92 &0.07$\pm$0.38 &7.59$\pm$7.39&15.19$\pm$14.78
&78.09 /72 &9&CPL\\\hline
&&&Main2:(0.141,0.262)&PL&&\textbf{1.01$\pm$0.10$^c$} &16.51$\pm$10.87&19.98$\pm$13.16&95.51/73 &CPL&127.76$\pm$108.59 &0.56$\pm$0.38 &-&-&93.12 /72 &2&PL\\\hline\hline
sGRBs+EE&&&&&&&&&&&&&&&\\ \hline
050724&98.684&0.257&Main:(-0.020,0.243)&PL&\emph{167.65$\pm1.19$}&\emph{1.41$\pm$0.08} &12.50$\pm$8.42&32.92$\pm$22.18&80.55/73 &CPL&179.46$\pm$185.48
&1.10$\pm$0.33&-&-& 79.50/72 &1&PL\\\hline
&&&EE:(0.922,1.247)&PL&\emph{52.64$\pm2.15$}&\emph{1.94$\pm$0.29} &1.56$\pm$0.79&5.08$\pm$2.58
&62.77/73 &CPL&120.22$\pm$329.47&1.56$\pm$1.15&-&-& 62.62 /72 &0&PL\\\hline
130603B&0.176&0.3565&Main1:(-0.008,0.051)&PL&&\textbf{0.62$\pm$0.06$^c$}&169.00$\pm$116.73&99.24$\pm$68.55&90.12/73&CPL&196.17$\pm$142.43&0.31$\pm$0.21&-&-&87.48/72&3&PL\\\hline
&&&Main2:(0.051,0.162)&PL&&\textbf{0.81$\pm$0.05$^c$}&74.00$\pm$51.20&82.44$\pm$57.04&70.41/73&CPL&
453.76$\pm$648.62&0.68$\pm$0.19&-&-&69.51/72&1&PL \\\hline
 &&&EE:(0.162,0.260)&PL&\emph{33.40$\pm0.93$}&\emph{2.20$\pm$0.15}&3.61$\pm$2.27&3.52$\pm$2.21&79.50/73&CPL&27.23$\pm$16.03&0.75$\pm$0.82&-&-&74.52/72&5&PL \\\hline
\end{tabular}
\begin{list}{}{}
\item $^a$ $E_p$ estimated using Eq. 5 for the segments whose best-fit model is PL with $1.2\leq\alpha\leq2.3$ and marked in italics \citep{Zhang2007a}. $^b$ The segments whose best-fit model is PL with $\alpha>2.3$ marked in boldface. $^c$ The segments whose best-fit model is PL with $\alpha<1.2$ marked in boldface. $^d$ $F_{lux}$ of the best-fit model in unit of $10^{-7} erg  cm^{-2} s^{-1}$. $^e$ $S_\gamma$ of the best-fit model in unit of $10^{-8} erg  cm^{-2} $.
\end{list}
\label{tab2}
\end{sidewaystable*}
\end{document}